\documentclass[showpacs,amsmath,amssymb,aps,prb,twocolumn]{revtex4-1}

\usepackage{graphicx}
\usepackage{dcolumn}
\usepackage{bm}

\usepackage{multirow}

\begin{document}

\title{What controls the temperature of a soft mode-driven structural
  phase transition?}

\author{Jacek C. Wojde{\l} and Jorge \'I\~niguez}

\affiliation{Institut de Ci\`encia de Materials de Barcelona
  (ICMAB-CSIC), Campus UAB, 08193 Bellaterra, Spain}

\begin{abstract}
We have used an effective model of ferroelectric PbTiO$_{3}$, which
displays a representative soft mode-driven phase transition, to
investigate how different features of the potential-energy surface
affect the transition temperature $T_{\rm C}$. We find that the energy
difference between PbTiO$_{3}$'s high-symmetry (cubic) and
low-symmetry (tetragonal) phases (which we call ground state energy
$E_{gs}$) is the parameter that most directly and strongly determines
$T_{\rm C}$. We have also found that other simple features of the
energy landscape, such as the amplitude of the distortion connecting
the high-symmetry and low-symmetry structures, can be used as a
predictor for $T_{\rm C}$ only as long as they are correlated with the
magnitude of $E_{gs}$. We discuss how our results relate to the
expected behaviors that can be derived from simpler theoretical
approaches, as well as to phenomenological studies in the
literature. Our findings support the empirical rule for estimating
$T_{\rm C}$ proposed by Abrahams {\sl et al}. [Physical Review {\bf
    172}, 551 (1968)] and clarify its physical interpretation. The
evidence also suggests that deviations from the expected behaviors are
indicative of complex lattice-dynamical effects involving strong
anharmonic interactions (and possibly competition) between the soft
phonon driving the transition and other modes of the material.
\end{abstract}





\maketitle

\section{Introduction}

Structural phase transitions driven by soft phonon
modes\cite{blinc-book1974,dove-book05} receive much attention for both
fundamental and technological reasons. The occurrence of a soft mode
is accompanied by a variety of striking effects, such as very large
responses (elastic, dielectric, piezoelectric) and highly tunable
properties, that can be exploited in applications. Hence, there is
interest in controlling the transition temperature $T_{\rm C}$, as
this will in turn determine the functional properties of the material
at specific (e.g., ambient) conditions. This interest is being
refueled by evidence that tuning the structural behavior provides us
with convenient strategies to enhance other important properties, such
as the magnetoelectric response.\cite{wojdel10}

From a designer's perspective, it would be useful to have simple rules
to estimate $T_{\rm C}$ from limited information about a compound. In
particular, if we were able to identify a simple {\em predictor} that
allowed us to guess $T_{\rm C}$ from routine first-principles
calculations, we could accelerate the discovery of materials that take
advantage of soft mode-related effects. Such a knowledge would also be
relevant to the construction of effective potentials for simulations
of lattice-dynamical phenomena, as it would tell us which key
properties the models must reproduce to render accurate $T_{\rm C}$'s.

\begin{figure*}[t!]
\includegraphics[width=0.95\linewidth]{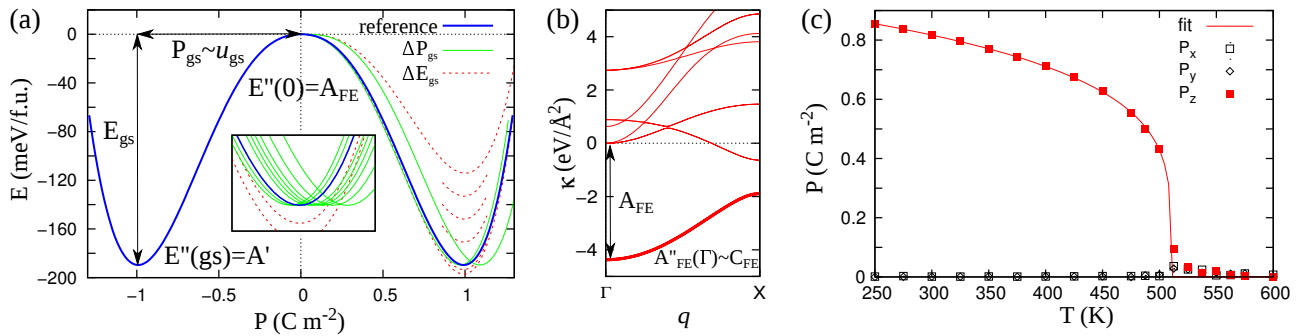}
\caption{(Color online) Panel~(a): The thick solid line represents the
  double well potential corresponding to the FE instability of our
  model for PTO. We also show energy wells corresponding to modified
  potentials, denoted ``(Ti--O)$^{4}$'' in the text, in which either
  the ground state polarization ($\Delta P_{gs} \neq 0$) or the ground
  state energy ($\Delta E_{gs} \neq 0$) has changed. The parameters
  characterizing the energy surface, and mentioned in the text, are
  indicated. Note that $A_{\rm FE} = \partial^{2}E/\partial P^{2}$
  evaluated at $P=0$, while $A' = \partial^{2}E/\partial P^{2}$
  evaluated at $P=P_{gs}$. Panel~(b): Harmonic force constants of the
  cubic phase of our PTO model, along the $\Gamma - X$ direction of
  the first Brillouin zone. We indicate the key parameters mentioned
  in the text. We use a thicker line to highlight the
  transversal-optical band corresponding to the FE instability. Note
  that $C_{\rm FE} = \partial^{2} A_{\rm FE}/\partial q^{2}$, for $q$
  along the $\Gamma - X$ line and evaluated at $\Gamma$. Panel~(c):
  Temperature dependence of the polarization as obtained from Monte
  Carlo simulations of our reference PTO model. The line corresponds
  to the fit to the model function described in the Appendix.}
\label{fig1}
\end{figure*}

The simplest atomistic model that captures the essence of a soft
mode-driven transition may be the so-called discrete $\phi^{4}$
model.\cite{bruce80} The potential energy is written as
\begin{equation}
E  = \sum_{i} \left[ \frac{A}{2} u_{i}^{2} + \frac{B}{4}
  u_{i}^{4} \right] + \frac{C}{2}\sum_{ii'}^{\rm n.n.} (u_{i}-u_{i'})^{2}
\, ,
\label{eq:model}
\end{equation}
which can be viewed as a Taylor series, around a reference structure
of zero energy, as a function of local structural distortions $u_{i}$
defined at every cell $i$. The collective condensation of these local
modes reduces the energy of the material according to a double-well
potential ($A<0$, $B>0$) like the one sketched in
Fig.~\ref{fig1}(a). Each local mode is coupled to its nearest
neighbors (n.n.) by a spring constant $C$ (here we take $C>0$) that
determines the dispersion of the associated phonon band [see
  Fig.~\ref{fig1}(b)]. Within this model, $E_{gs} = - A^{2}/4B$ is the
energy {\sl per} cell of the ground state structure, which is
characterized by $u_{i} = u_{gs} = \sqrt{-A/B}$ $\forall i$. This
low-energy phase is reached from the high-symmetry structure ($\langle
u_{i} \rangle = 0$ $\forall i$, where $\langle ... \rangle$ denotes
thermal average) when we bring the system below $T_{\rm C}$. Note that
$|E_{gs}|$ roughly quantifies the thermal energy that the system needs
to {\sl jump} between equivalent potential wells and thus stabilize
the high-symmetry phase. Hence, it is tempting to assume
\begin{equation}
k_{\rm B}T_{\rm C} \sim |E_{gs}| = A^{2}/4B  \, ,
\label{eq:energy}
\end{equation}
where $k_{\rm B}$ is Boltzmann's constant. Since the calculation of
$E_{gs}$ from first-principles is a trivial task, this would be a very
convenient $T_{\rm C}$ predictor.

However, a more careful analysis suggests that the above choice might
not be optimal. Within the mean-field approximation,\cite{bruce80} it
is possible to solve the $\phi^{4}$ model in the displacive ($|A|/C\ll
1$) and order-disorder ($|A|/C\gg 1$) limits (i.e., for strongly- and
weakly-coupled local modes, respectively). In both cases we get
\begin{equation}
k_{\rm B}T_{\rm C} \sim C u_{gs}^{2} = C|A|/B \, .
\label{eq:mean-field}
\end{equation}
This predictor gathers information about the magnitude of the
structural instability (quantified by $u_{gs}^{2}$ instead of
$|E_{gs}|$) and the energy cost for the occurrence of alternative,
inhomogeneous distortions (given by $C$).

Finally, it has been found empirically\cite{abrahams68} that $T_{\rm
  C}$ correlates with the magnitude of the symmetry-breaking
distortion, so that
\begin{equation}
T_{\rm C} \sim u_{gs}^{n} \, ,
\label{eq:empirical}
\end{equation}
where $n$ is a positive integer. By examining the structural phase
transitions of a variety of ferroelectric compounds, the authors of
Ref.~\onlinecite{abrahams68} concluded that, to fit their data, $n$
can be chosen to be either 1 or 2. Yet, they argue that $n = 2$
renders a physically sounder relation, an interpretation that was
backed shortly after by the theoretical work of
Lines.\cite{lines69,lines-book1977} It has been shown more
recently\cite{grinberg04,juhas04} that Eq.~(\ref{eq:empirical}) with
$n = 2$ renders a good description for the ordering temperatures of a
family of ferroelectric relaxor perovskites.

The above mentioned {\sl laws} have intriguing implications. For
example, the validity of Eq.~(\ref{eq:empirical}) suggests that either
the mean-field result of Eq.~(\ref{eq:mean-field}) is not realistic or
that the parameter $C$ adopts similar values in all the materials that
were investigated in
Refs.~\onlinecite{abrahams68,grinberg04,juhas04}. Also, the validity
of Eqs.~(\ref{eq:energy}) or (\ref{eq:empirical}) might imply that
$T_{\rm C}$ does not significantly depend on the energetics of
distortions not present in the ground state. Further,
Eqs.~(\ref{eq:mean-field}) and (\ref{eq:empirical}) suggest that one
may encounter materials with very strong instabilities, even with
$|E_{gs}| \gg Cu_{gs}^{2}$, that might nevertheless display a
relatively low $T_{\rm C}$ determined by relatively small values of
$u_{gs}$ and $C$. These are all rather surprising notions.

To shed light on these issues, we conducted a series of numerical
experiments using a model potential for PbTiO$_{3}$ (PTO). PTO
presents a prototypic structural transition, between high-temperature
cubic and low-temperature tetragonal structures, and is representative
of the class of materials for which empirical rules like
Eq.~(\ref{eq:empirical}) have been observed to hold. The employed
model describes PTO in full atomistic detail, and its parameters can
be modified by hand to study the resulting changes in $T_{\rm C}$. We
can thus test the performance of the predictors mentioned above.

\section{Computational experiments}

Our model for PTO is described in Ref.~\onlinecite{wojdel13}, where it
is labeled ``$L^{I}$''. It can be viewed as a Taylor series of the
energy, around the ideal cubic perovskite structure, as a function of
all possible atomic distortions and strains. The series was truncated
at 4th order and only pairwise interaction terms were included.
Hence, in essence, our PTO model can be seen as an extended version of
the $\phi^{4}$ Hamiltonian in which all the degrees of freedom are
treated explicitly. The potential parameters were computed by using
the local density approximation (LDA) to density functional theory. To
compensate for LDA's well-known overbinding problem, we simulate the
model under the action of a tensile hydrostatic pressure of 14.9~GPa.

The potential well associated with the ferroelectric (FE) instability
of our model for PTO is shown in Fig.~\ref{fig1}(a). When we solve the
model by running Monte Carlo (MC) simulations in a periodically
repeated box of 10$\times$10$\times$10 unit cells, we obtain an abrupt
transition at $T_{\rm C}\approx$~510~K,\cite{fnTC-PTO} as reflected in
the $T$-dependence of the polarization ($P$) in Fig.~\ref{fig1}(c).
In order to get reliable results for atomic displacements and strains
(from which we derive the spontaneous polarization as described in
Ref.~\onlinecite{wojdel13}), we ran at least 20,000~MC sweeps for
thermalization, followed by at least 20,000 additional sweeps to
compute thermal averages. For temperatures close to the transition,
the simulations were run for up to 80,000~MC sweeps after
thermalization in order to obtain well converged values. We
initialized all our simulations, for all models and temperatures, from
the same cubic reference state; hence, our results do not display any
hysteretic behavior. For the original model, we also ran simulations
in which, for each new temperature $T + \Delta T$, we used a
representative configuration of the previous temperature considered
$T$ to initialized the MC simulations; we found that the hysteresis,
if present, is narrower than what we claim for the accuracy of the
$T_{\rm C}$ determination.\cite{fn-hysteresis} From the obtained
thermal averages, we estimate $T_{\rm C}$ by a simple and robust
fitting to the $P(T)$ profile, as described in the Appendix.

\begin{figure}[b!]
\includegraphics[width=0.90\linewidth]{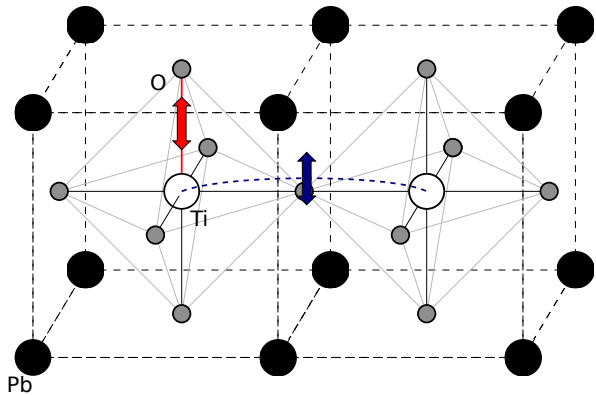}
\caption{Sketch of the Ti--O and Ti--Ti interactions that we modified
  in our models. We worked in the displacement-difference
  representation introduced in Ref.~\protect\onlinecite{wojdel13}. In
  the Ti--O case, the interaction involves displacements along the
  direction defined by the pair of atoms involved; it can be easily
  seen that they do not affect the energetics of the O$_{6}$
  octahderal rotations discussed in Section~III. In the Ti--Ti case,
  the interaction involves displacements orthogonal to the direction
  defined by the pair of atoms involved. The modified interaction
  controls the transversal modulation of the FE instability as we move
  away from the center of the Brillouin zone.}
\label{fig2}
\end{figure}

We began by checking how the separate variations of the ground state
energy $|E_{gs}|$ and polarization $P_{gs}$ affect $T_{\rm C}$. To do
so, we constructed models whose associated energy wells are shown in
Fig.~\ref{fig1}(a). Such models were obtained by tuning some of the
interactions controlling the FE instability, namely, the harmonic and
4th-order couplings between neighboring Ti and O atoms (see sketch in
Fig.~\ref{fig2}). We were thus able to (1) change $E_{gs}$ while
keeping $P_{gs} = P_{gs}^{0}$, where the ``0'' superscript denotes
values corresponding to our unmodified PTO model, and (2) shift
$P_{gs}$ while keeping $E_{gs} = E_{gs}^{0}$. (We also worked with
Pb--O couplings and obtained very similar results for the behavior of
$T_{\rm C}$.)  All the modified models we studied present the same
qualitative behavior, and the atomic distortions characterizing the FE
phase resemble closely those occurring in real PTO. Note that large
changes in the potential parameters can eventually lead to
qualitatively different behaviors (e.g., suppression of
ferroelectricity, change of polar axis), which limited our ability to
tune the models. Finally, let us mention that the potential parameters
we modified do not interfere with the energetics of the PTO modes
involving rotations of the oxygen octahedra; hence, our changes did
not affect significantly the instability competition discussed below.

Note that we decided to use $P_{gs}$ as a measure of the total
distortion $u_{gs}$. Besides its historical
motivation,\cite{abrahams68} this choice is reasonable because in PTO
all the individual atomic displacements, as well as the cell strain,
add up to the total polarization of the ground state. (See caption of
Table~I for some detail on how these quantities are connected.) At any
rate, we checked that our qualitative conclusions remain the same if
the bare atomic displacements, instead of the associated polarization,
are considered.

\begin{table}[b!]
\caption{Computed Curie temperatures ($T_{\rm C}$, given in Kelvin)
  for a few representative models considered in this work. The models
  are labeled by indicating which specific couplings, and up to which
  highest order, were modified. For each model, we give a number of
  key parameters that characterize the potential energy surface and
  are described in the text. The reference values of the parameters
  are $E_{gs}^{0} = -$190~meV/f.u., $P_{gs}^{0} = 0.99$~C/m$^{2}$,
  $C_{\rm FE}^{0} = $14.66~eV/\AA$^{2}$, $A_{\rm FE}^{0} =
  -$4.37~eV/\AA$^{2}$, and $A'^{0} = 10.23$~eV/\AA$^{2}$. Note that,
  if we give the curvatures $A_{\rm FE}$ and $A'$ in force-constant
  units, $A_{\rm FE}$ coincides exactly with the stiffness of the FE
  instability of the cubic phase [see $\kappa$ bands in
    Fig.~\protect\ref{fig1}(b)]. Alternatively, we have $A_{\rm
    FE}^{0} = -$0.744~eV~m$^{4}$/C$^{2}$ and $A'^{0}
  =1.783$~eV~m$^{4}$/C$^{2}$; in essence, the transformation from
  atomic distortion (given in \AA) to polarization (given in
  C/m$^{2}$) involves the unit cell volume and the polarity of the FE
  instability, which take values of about 64~\AA$^{3}$ and 10
  elemental charges, respectively, in our case. Note also that $C_{\rm
    FE} = \partial^{2}A_{\rm FE}/\partial q^{2}$, where we assume $q$
  is given with respect to the reciprocal lattice vectors, and is thus
  adimensional. }
\vskip 1mm

\begin{tabular}{ccccccc}
\hline\hline 
\rule{0pt}{3ex}model & $E_{gs}/E_{gs}^{0}$ & $P_{gs}/P_{gs}^{0}$ &
$C_{\rm FE}/C_{\rm FE}^{0}$ & $A_{\rm FE}/A_{\rm FE}^{0}$ & $A'/A'^{0}$ &
$T_{\rm C}$ \\  \hline
\rule{0pt}{3ex}original & 1 & 1 & 1 & 1 & 1 & 510 \\[1ex]
\multirow{4}{*}{(Ti--O)$^{4}$} 
& 0.60 & 1    & 0.79 & 0.96 & 0.30 & 373 \\
& 1.04 & 1    & 1.02 & 1.02 & 1.05 & 521 \\
& 1    & 0.98 & 1.02 & 1.02 & 1.05 & 510 \\
& 1    & 1.10 & 0.92 & 0.98 & 0.77 & 475 \\[1ex]
\multirow{2}{*}{(Ti--Ti)$^{2}$}
& 1    & 1    & 0.93 & 1    & 1    & 490 \\
& 1    & 1    & 3.39 & 1    & 1    & 732 \\[0.5ex]
\multirow{4}{*}{(Ti--O)$^{8}$}
& 0.83 & 1    & 1    & 1    & 0.65 & 463 \\
& 1.10 & 1    & 1    & 1    & 1.51 & 538 \\
& 1    & 0.91 & 1    & 1    & 1.35 & 520 \\
& 1    & 1.11 & 1    & 1    & 0.31 & 490 \\
\hline\hline
\end{tabular}
\label{table}
\end{table}

\begin{figure}[t]
\includegraphics[width=0.95\linewidth]{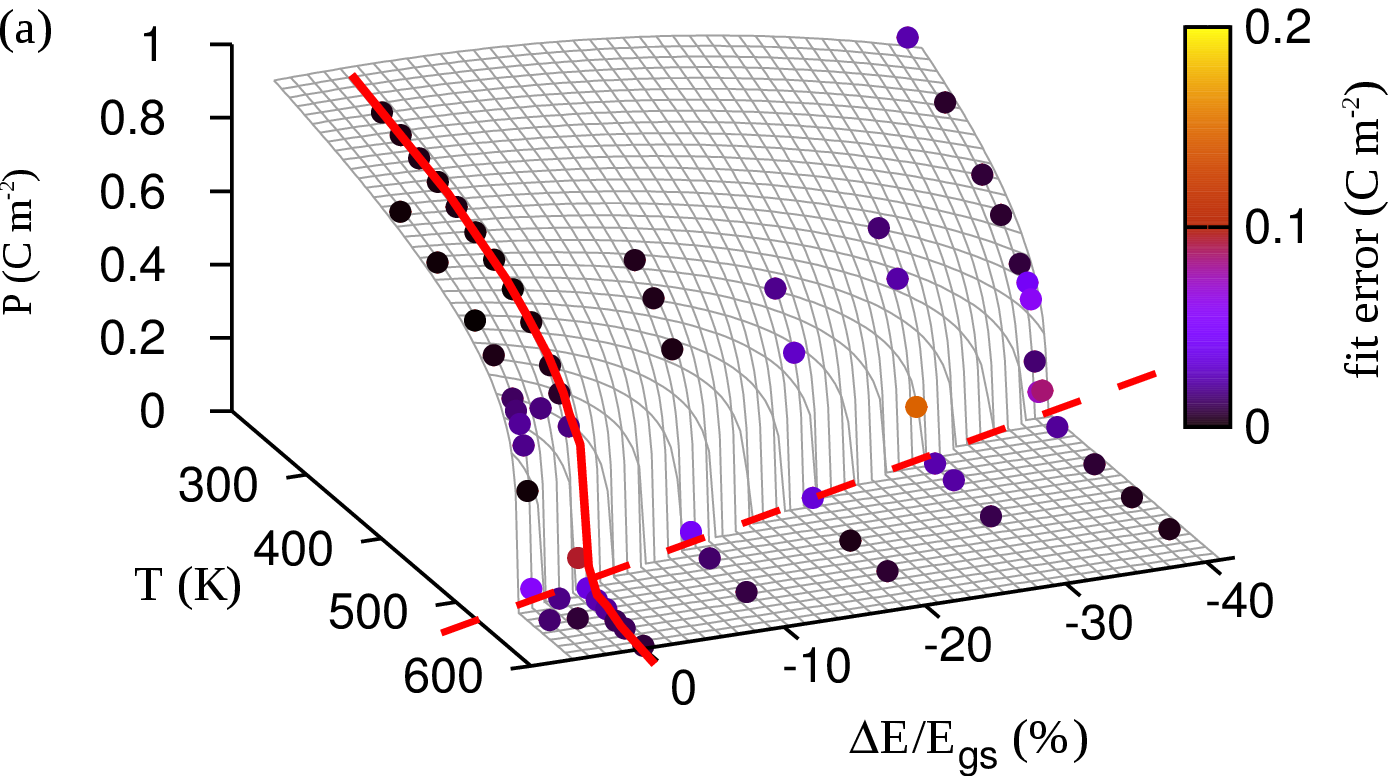}
\includegraphics[width=0.95\linewidth]{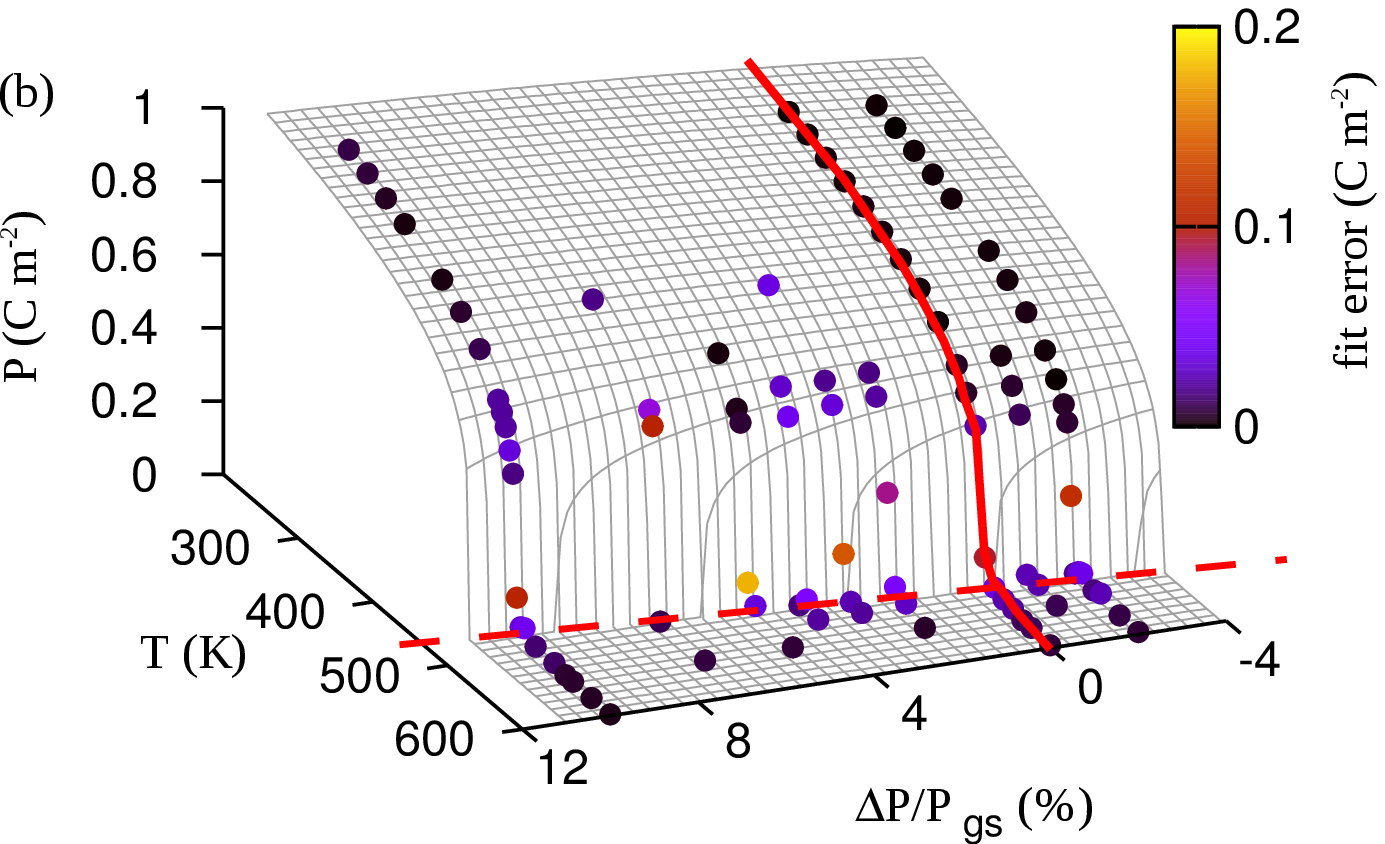}
\caption{(Color online) Computed $P(T)$ curves for a variety of
  ``(Ti--O)$^{4}$'' modified models. Panel~(a): Models in which
  $E_{gs}$ changes while $P_{gs} = P_{gs}^{0}$ is kept
  constant. Panel~(b): Models in which $P_{gs}$ changes while $E_{gs}
  = E_{gs}^{0}$ is kept constant.}
\label{fig3}
\end{figure}

Figure~\ref{fig3}(a) shows the results obtained when we varied
$E_{gs}$ at constant $P_{gs}^{0}$. Clearly, modifying $E_{gs}$ can
lead to large shifts in $T_{\rm C}$ (e.g., $T_{\rm C}$ decreases by
about 137~K when $|E_{gs}|$ is 40\% smaller), and the dependence is
approximately linear. Hence, these results support the heuristic
assumption that $|E_{gs}|$ is a good predictor for $T_{\rm C}$. On the
other hand, Fig.~\ref{fig3}(b) shows the results obtained when we
varied $P_{gs}$ at constant $E_{gs}^{0}$. The range of $P_{gs}$ values
that we can explore is somewhat limited, yet sufficient to observe a
surprising effect: Increasing $P_{gs}$ leads to a reduction of $T_{\rm
  C}$. This is in obvious disagreement with the mean-field
[Eq.~(\ref{eq:mean-field})] and empirical [Eq.~(\ref{eq:empirical})]
expectations mentioned above.

This apparent failure of the mean-field prediction is shocking, as
previous works on related models suggest that such an approximation
should be able to capture the main qualitative behaviors of our PTO
potential.\cite{lines69,pytte69,waghmare97} Let us consider in some
detail such a discrepancy. Manipulating the 2nd- and 4th-order Ti--O
couplings in our PTO model may seem equivalent to tuning the
parameters $A$ and $B$ in $\phi^{4}$ Hamiltonian
[Eq.~(\ref{eq:model})]. An important difference, though, is that such
a modification of our potential involves changes in the dispersion of
the phonon bands. For example, Table~\ref{table} gives information
about a couple of constant-$E_{gs}^{0}$ models that present different
$P_{gs}$ values [see potentials labeled ``(Ti--O)$^{4}$'', where the
  notation indicates the highest-order coupling that was modified to
  construct them]. The $C_{\rm FE}$ parameter given in the Table
quantifies the curvature at $\Gamma$ of the bands associated with the
FE instability [see Fig.~\ref{fig1}(b)], and is analogous to the $C$
parameter of the $\phi^{4}$ Hamiltonian. Interestingly, in our
constant-$E_{gs}^{0}$ models, larger $P_{gs}$ values correspond to
smaller $C_{\rm FE}$ curvatures, and smaller $C_{\rm FE}$'s are
consistent with the observed decrease in $T_{\rm C}$ according to
Eq.~(\ref{eq:mean-field}). Thus, a reduction in $T_{\rm C}$ for
increasing $P_{gs}$ does not necessarily imply the failure of
Eq.~(\ref{eq:mean-field}).

We were able to specifically confirm the influence of $C_{\rm FE}$ on
the obtained $T_{\rm C}$'s. To do so, we constructed models in which
$C_{\rm FE}$ was modified while keeping $E_{gs} = E_{gs}^{0}$ and
$P_{gs} = P_{gs}^{0}$ constant, which required the introduction of an
additional harmonic coupling between neighboring Ti atoms (see sketch
in Fig.~\ref{fig2}). Table~\ref{table} shows the results for two
representative cases, labeled ``(Ti--Ti)$^{2}$''. The observed
behavior makes good physical sense: A larger $C_{\rm FE}$ implies a
greater energy cost for the occurrence of inhomogeneous locally-polar
distortions that are mutually exclusive with the dominant FE soft
mode, and hence results in a higher $T_{\rm C}$. Additionally, we can
numerically evaluate Eq.~(\ref{eq:mean-field}) using the information
in Table~\ref{table} for the ``(Ti--O)$^{4}$'' models with constant
$E_{gs}^{0}$. Thus, for example, Eq.~(\ref{eq:mean-field}) predicts
that our model with $P_{gs}/P^{0}_{gs}$~=~1.10 and $C_{\rm FE}/C_{\rm
  FE}^{0}$~=~0.92 should present an enhancement of about 11\% in
$T_{\rm C}$; however, such a prediction is in obvious disagreement
with the computed decrease.

Interestingly, it may seem that our data suggest an alternative
predictor for $T_{\rm C}$. For the models in Table~\ref{table}, we
report the curvature of the $E(P)$ curve at $P=0$ [$E''(0)$ or $A_{\rm
    FE}$ in Fig.~\ref{fig1}(a)], which essentially corresponds to the
parameter $A$ of the $\phi^{4}$ Hamiltonian. This parameter measures
how unstable the $P=0$ paraelectric (PE) state is: large negative
values of $A_{\rm FE}$ imply a greater difficulty to stabilize the PE
phase, and should thus correspond to higher $T_{\rm C}$'s. Hence, one
may heuristically propose $T_{\rm C} \sim |A_{\rm FE}|$, which is
essentially satisfied by all the models we studied. It is worth noting
that, in the case of the ``(Ti--O)$^{4}$'' models in which we vary the
ground state polarization at constant $E_{gs}^{0}$, changes in
$P_{gs}$ and $|A_{\rm FE}|$ are forcefully correlated, an increase of
the former implying a decrease of the latter. Hence, the $T_{\rm C}
\sim |A_{\rm FE}|$ and $T_{\rm C} \sim P_{gs}^{2}$ rules are
incompatible in this case, and we find that the former matches our
Monte Carlo results better. We further investigated the validity of
this new rule by considering other modified models (not shown here).
Ultimately, we found that, in the constant-$E_{gs}^{0}$
``(Ti--O)$^{4}$'' cases of Table~\ref{table}, we should attribute the
changes in transition temperature to the variations in the $C_{\rm
  FE}$ parameter rather than to changes in $A_{\rm FE}$. Nevertheless,
we did obtain additional indications of the importance of the details
of the $E(P)$ curve from our last set of modified models, which we
describe in the following.

\begin{figure}
\includegraphics[width=0.95\linewidth]{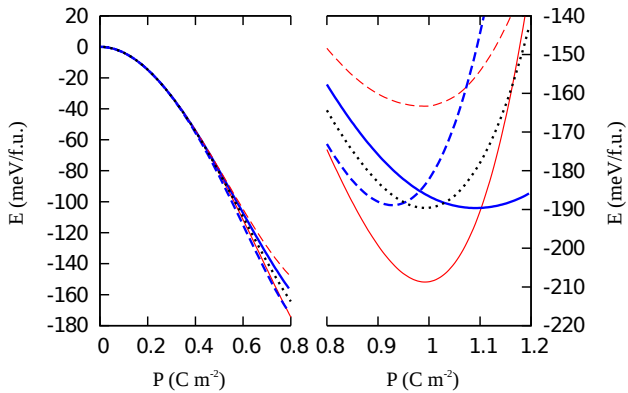}
\caption{(Color online) Comparison of the double-well potential for
  the FE instability of PTO corresponding to our reference model
  (black dotted line) and the ``(Ti--O)$^{8}$'' modified potentials
  described in the text. Red and blue lines correspond to
  constant-$P_{gs}$ and constant-$E_{gs}$ models, respectively. These
  $E(P)$ curves were obtained by determining, for each $P$ value, the
  structure that minimizes the energy and can be described with a
  5-atom unit cell.}
\label{fig4}
\end{figure}

Finally, we constructed models in which only $E_{gs}$ or $P_{gs}$ vary
while the other parameters discussed so far ($C_{\rm FE}$ and $A_{\rm
  FE}$) are kept fixed. Doing this required the tuning of Ti--O
couplings up to 8th-order while keeping the harmonic interactions
constant; these models are labeled ``(Ti--O)$^{8}$'' in
Table~\ref{table}. Our results ratify that $E_{gs}$ has a considerable
impact on $T_{\rm C}$, in qualitative agreement with
Eq.~(\ref{eq:energy}). We also find that varying $P_{gs}$ alone has an
effect on $T_{\rm C}$. However, once again we observe that a larger
ground state distortion leads to a smaller transition temperature, in
disagreement with initial expectations.

When we examined this last set of modified models, we observed that a
larger $P_{gs}$ value corresponds to a shallower energy surface around
the FE minimum (see Fig.~\ref{fig4}). In view of this, we reexamined
all our potentials and found that the curvature at $P = P_{gs}$, which
we call $A'$ in Table~I, does correlate with the computed transition
temperatures. We might thus speculate that, generally speaking, {\em
  stiffer} FE phases (with larger $A'$ values associated to them) will
be more difficult to destabilize, and will thus correspond to higher
$T_{\rm C}$'s. This observation suggests yet another heuristic
predictor for the transition temperature, namely, $T_{\rm C} \sim
A'$. It is interesting to note that for the $\phi^{4}$ potential we
have $A' = -2 A$; hence, in this case we can write $T_{\rm C} \sim A'
\sim |A|$, which coincides with the above mentioned predictor based on
curvature of $E(P)$ around $P=0$.

A natural next step would be to investigate models in which we would
keep all the parameters in Table~I constant except for $E_{gs}$ or
$P_{gs}$. This would require our introducing couplings of even higher
order, and producing ever more artificial potentials. Hence, we did
not pursue this line any further.

\section{Discussion}

In view of these results, what is the status of the predictors for
$T_{\rm C}$ mentioned in the introduction? Are our findings compatible
with existing literature? What are the lessons to be learned? 

\subsection{Implications for $T_{\rm C}$ predictors}

We have found that the mean-field result for the $\phi^{4}$ model
[Eq.~(\ref{eq:mean-field})], which is essentially equivalent to the
formulas proposed by a number of authors,\cite{lines69,pytte69} does
not give an accurate description of the behavior of our simulated
materials. The main conflict concerns the qualitative dependence of
$T_{\rm C}$ on $P_{gs}^{2}$, as we have found that our simulations
render a behavior ($T_{\rm C}$ decreases with growing $P_{gs}^{2}$)
that is just opposed to the expected one. This is a serious
discrepancy, as the $\phi^{4}$ model can be seen as a simplified
version of our PTO potential, and we certainly expect its behavior to
be qualitatively well captured by the mean-field approximation. Hence,
how can we explain this apparent contradiction?

Let us begin by noting that, while we are in principle entitled to
associate parameters in the $\phi^{4}$ model with the analogous
quantities for our PTO potential, this correspondence is not a strict
one. The $\phi^{4}$ model has only three independent constants --
which we can choose to be $A$, $B$, and $C$ in Eq.~(\ref{eq:model}) --
and, once those are given, we can write simple relationships between
the derived quantities. Thus, for example, the equality $u_{gs}^{2} =
4 E_{gs}/A$ is always fulfilled by the $\phi^{4}$ potential. However,
as one can easily check from the information in Table~\ref{table}, the
analogous identity $P_{gs}^{2} = \gamma E_{gs}/A_{\rm FE}$, where
$\gamma$ is an appropriate constant, does not hold for our PTO
potentials. For example, for the model described in the second line of
Table~\ref{table} we have $(P_{gs}/P_{gs}^{0})^{2} = 1$, which differs
a lot from $E_{gs}/E_{gs}^{0} \times A_{\rm FE}^{0}/A_{\rm FE} =
0.63$. It is thus obvious that the relationships that are valid for
the $\phi^{4}$ model, even the simplest ones pertaining to the ground
state properties and potential shape, do not necessarily apply to more
realistic models of structural transitions.

The reason for such differences lies on the inherent complexity of our
reference PTO potential. Indeed, even for a 4th-order model like ours,
the energy well corresponding to the FE instability can effectively be
of a higher polynomial order due to the anharmonic couplings between
the unstable FE mode and other modes and strains in the material. This
is a critical difference with respect to the $\phi^{4}$ model, and the
likely cause of the discrepancies mentioned above. To understand this
better, consider the energy of a simple model
\begin{equation}
E = a u^{2} + b u^{4} + c v^{2} + d v^{4} + e v u^{3} \, ,
\label{eq:E-model}
\end{equation}
where $u$ is the amplitude of a soft mode with an associated
double-well potential ($a<0$, $b>0$) and $v$ is the amplitude of a
stable mode ($c,d>0$) coupled with $u$ anharmonically. In such a case,
it is easy to see that, upon condensation of $u \neq 0$, we get a
secondary distortion
\begin{equation}
v = - \frac{e}{2c} u^{3} \, ,
\label{eq:secondary-mode}
\end{equation}
where, for simplicity, to derive this expression we have assumed that
$v$ will always be relatively small. By substituting
Eq.~(\ref{eq:secondary-mode}) into Eq.~(\ref{eq:E-model}) we get a
renormalized energy for $u$ that reads
\begin{equation}
E' = a u^{2} + b u^{4} - \frac{e^{2}}{4c} u^{6} + \frac{d e^{4}}{16c^{4}}
u^{12} \, ,
\end{equation}
which is effectively of 12th order. [The ${\cal O}(v^{4})$ term that
  eventually leads to the ${\cal O}(u^{12})$ contribution has been
  included here to make it clear that the renormalized energy
  continues to be bounded from below, despite the ${\cal O}(u^{6})$
  term being negative.]  This example constitutes a realistic
description of how the FE soft mode ($u$) and other $\Gamma$-point
modes with the same polar symmetry ($v$) interact anharmonically. Such
couplings are responsible for a variety of effects in PTO, such as the
significant differences that exist between the atomic displacements
corresponding to the FE ground state and those associated with the
eigenvector of the FE soft mode of the cubic phase, etc. The precise
list of anharmonic couplings in our PTO potential can be found in
Table~1 of Ref.~\onlinecite{wojdel13}.

It is clear that our model for PTO is closer to reality than the
simple $\phi^{4}$ Hamiltonian. Hence, one may wonder: do our results
imply that the $T_{\rm C} \sim C u_{gs}^{2}$ rule should not be able
to describe real materials? To answer this question, let us note that
the modified models considered in this work correspond to somewhat
artificial situations in which changes in $P_{gs}$ and $E_{gs}$ are
strictly decoupled. In contrast, in reality one expects strong soft
mode instabilities to involve large values of {\sl both} $|E_{gs}|$
and $P_{gs}$. (This is indeed what we obtain for our PTO model when we
simply crank up the magnitude of the interactions responsible for the
FE instability without imposing any constraint.)  Hence, leaving aside
unusual choices of potential parameters, we can generally expect
$|u_{gs}|^{n} \sim |E_{gs}|^{m} \sim s$, where $s$ measures the {\sl
  strength} of the structural instability, and $n$ and $m$ are two
positive numbers. In other words, in general we expect $u_{gs}^{2}$
and $|E_{gs}|$ to be essentially equivalent as $T_{\rm C}$ predictors.

Now, even though they may correspond to unusual situations, the
unexpected results that we obtained for our modified models with
constant-$E_{gs}^{0}$ do suggest some relevant conclusions. As
mentioned above, when subject to the constant-$E_{gs}^{0}$ constraint,
an increase of $P_{gs}$ involves changes in the potential surface
(e.g., around $P = 0$ and/or $P = P_{gs}$) that tend to result in a
lower $T_{\rm C}$. This finding indicates that: (1) Subtle changes in
the potential surface can have an important impact on the computed
transition temperature. As far as we know, this is an effect that had
not been noticed before, and one that is important to keep in mind if
we want to construct first-principles model potentials (like those of
Ref.~\onlinecite{wojdel13} and
others\cite{zhong94a,shin05,sepliarsky05}) that render accurate
transition temperatures. (2) The fundamental quantities for predicting
$T_{\rm C}$ are those directly related to the energy. Thus, our
results suggest that $P_{gs}^{2}$ or $u_{gs}^{2}$ may act as
predictors only because their magnitude is usually connected with the
strength of the structural instability. In contrast, $|E_{gs}|$ is a
fundamentally more robust predictor for $T_{\rm C}$.

Of course, the above arguments imply that, in our opinion, our results
are perfectly compatible with the empirical rule $T_{\rm C} \sim
u_{gs}^{2}$ proposed by Abrahams {\sl et al}.\cite{abrahams68} In
fact, it is worth mentioning that, when discussing the physical
interpretation of their newly-found law, these authors viewed the
value of $u_{gs}^{2}$ as a measure of the ``thermal energy at the
Curie point''. We believe that such an interpretation is the most
natural one, and it falls in line with our conclusions.

Finally, let us note that we numerically fitted our results for
$T_{\rm C}$ using an expression of the form
\begin{equation}
\begin{split}
\frac{T_{\rm C}}{T_{\rm C}^{0}} = &
\left( \frac{E_{gs}}{E_{gs}^{0}} \right)^{n_1} \times 
\left( \frac{P_{gs}}{P_{gs}^{0}} \right)^{n_2} \times
\left( \frac{C_{\rm FE}}{C_{\rm FE}^{0}} \right)^{n_3} \\
& \times \left( \frac{A_{\rm FE}}{A_{\rm FE}^{0}} \right)^{n_4} \times
\left( \frac{A'}{A'^{0}} \right)^{n_5} \, ,
\end{split}
\end{equation}
where the exponents are adjustable real numbers. This exercise showed
that the $A_{\rm FE}$ and $A'$ parameters are comparatively
unimportant for determining $T_{\rm C}$, and can be neglected in a
first approximation. Further, we obtained $n_{1} = 0.49$ for $E_{gs}$,
$n_{2} = -0.34$ for $P_{gs}$, and $n_{3} = 0.29$ for $C_{\rm FE}$,
reflecting the dominant role of the energy difference between the
high- and low-symmetry structures. As expected from our results
summarized in Table~I, the magnitude of $P_{gs}$ is found to be
inversely proportional to $T_{\rm C}$. At the same time, in typical
cases in which $E_{gs}$ and $P_{gs}$ are correlated and grow/decrease
together, we can expect the $E_{gs}/E_{gs}^{0}$ term to dominate.

\subsection{Non-trivial cases: competing instabilities}

Let us now comment on the related works of Grinberg and
Rappe\cite{grinberg04} and Juhas {\sl et al}.,\cite{juhas04} who
combined first-principles results and experimental information to
empirically identify predictors for $T_{\rm C}$. These authors studied
a number of complex solid solutions involving PbTiO$_{3}$ and
PbZrO$_{3}$ crystals mixed with partly-disordered perovskites
PbMg$_{1/3}$Nb$_{2/3}$O$_{3}$, PbZn$_{1/3}$Nb$_{2/3}$O$_{3}$, and
PbSc$_{2/3}$W$_{1/3}$O$_{3}$. Experimentally, these materials are
found to behave as relaxor ferroelectrics, with the temperature
$T_{\epsilon,{\rm max}}$ -- corresponding to the maximum dielectric
response -- being the closest analogue of the Curie point of a normal
ferroelectric. By comparing the experimental values for $T_{\epsilon,
  {\rm max}}$ with the computed $u_{gs}$ ($u_{gs}$ quantifies a local
symmetry-breaking distortion in this case), a good correlation of the
form $T_{\epsilon, {\rm max}} \sim u_{gs}^{2}$ was observed. At first
sight, this finding seems to ratify the conclusions of Abrahams {\sl
  et al}.,\cite{abrahams68} and seems perfectly compatible with the
above discussion of our results. However, the authors of
Ref.~\onlinecite{juhas04} also observed clear deviations from the
$T_{\epsilon, {\rm max}} \sim |E_{gs}|$ rule that would be analogous
to Eq.~(\ref{eq:energy}). How does this affect our conclusion that
$|E_{gs}|$ is the more fundamental and robust predictor for $T_{\rm
  C}$?

A careful inspection of the data in Ref.~\onlinecite{juhas04} (see
e.g.  Table~IV in that paper) suggests that there are subtleties
hiding behind the proposed $T_{\epsilon, {\rm max}} \sim u_{gs}^{2}$
rule. As expected for Pb-based ferroelectrics and relaxors, Juhas {\sl
  et al.}  found that the distortions characterizing the low-symmetry
phases are dominated by the off-centering of the Pb atoms, with the
displacements of the $B$-site cations (Ti, Zr, etc.) being smaller by
a factor of 2 or 3, typically. However, their results also show that
the magnitude of the Pb displacements does {\sl not} correlate well
with $T_{\epsilon, {\rm max}}$. Instead, their data suggest that what
correlates strongly with $T_{\epsilon, {\rm max}}$ is the displacement
of the $B$-site cations, and it is such a correlation what ultimately
justifies the $T_{\epsilon, {\rm max}} \sim u_{gs}^{2}$ rule. (The
precise relationship was obtained as the result of a fitting procedure
in which the contributions from the Pb and $B$-atom displacements were
considered separately.\cite{juhas04}) Such an atomistic foundation
for the $u_{gs}^{2}$ predictor, with the relatively small
displacements of the $B$-cations dominating the effect, is truly
intriguing.

Interestingly, when discussing their $T_{\epsilon, {\rm max}} \sim
u_{gs}^{2}$ rule, Juhas {\sl et al}. wrote that the change in
$T_{\epsilon, {\rm max}}$ is not ``directly caused by changes in the
structural features such as the cation shifts, but is due to the
changes in the energetics of competing instabilities''. This is a
subtle point that is worth discussing. The competing instabilities
mentioned by these authors are the local polar distortions (which
ultimately prevail) and the so-called {\sl anti-ferrodistortive} (AFD)
modes involving concerted rotations of the oxygen octahedra in the
perovskite structure. We have recent and clear evidence that this kind
of competition has a large effect on the Curie temperature of
PTO\cite{wojdel13} and related materials.\cite{kornev06} In
particular, as shown in Ref.~\onlinecite{wojdel13} for the case of our
PTO model, it is possible to modify the energetics of the
oxygen-octahedra rotations (e.g., artificially suppressing them) and
obtain an effect in $T_{\rm C}$ (a very large increase), even if all
the key parameters describing the FE instability and ground state
($E_{gs}$, $P_{gs}$, $C_{\rm FE}$, $A_{\rm FE}$, and $A'$) remain
constant. Hence, {\sl a priori} there is no reason to expect the above
mentioned $T_{\rm C}$ predictors to describe well the behavior of
materials in which this type of {\sl hidden} effects are important. In
fact, it seems natural to suspect that departures from the {\sl
  normal} behavior may indicate the presence of this kind of
phenomena.

One can thus conjecture that, in the cases considered in
Refs.~\onlinecite{grinberg04} and \onlinecite{juhas04}, larger
displacements of the $B$-site cations probably correspond to weaker
AFD instabilities, which would in turn result in a less important
competition and a higher $T_{\epsilon,{\rm max}}$. Note that this
connection is not inconsistent with the observation that, in the
AFD-dominated phases of many perovskite oxides, the $B$-site cations
usually stay at the center of O$_{6}$ octahedra. This tendency can be
explained by the size effects captured by the so-called {\sl tolerance
  factor}.\cite{davies08,fornari01,reaney94}

It thus seems that the simple-looking predictor for $T_{\epsilon, {\rm
    max}}$ proposed in Refs.~\onlinecite{grinberg04,juhas04} hides
rather complex structural and lattice-dynamical mechanisms behind
it. As just mentioned, in these Pb-based relaxors $u_{gs}^{2}$ seems
to (anti)correlate with the importance of the FE--AFD competition,
which in turn controls the ordering temperature. In contrast, due to
the structural complexity of these materials, $u_{gs}^{2}$ does not
correlate well with the depth of the potential energy wells. In view
of this, the results of Refs.~\onlinecite{grinberg04,juhas04} cannot
be taken as support for the conclusions of Abrahams {\sl et
  al}.,\cite{abrahams68} which rely on the connection between $u_{gs}$
and $E_{gs}$ as emphasized above. Nevertheless, such results do show
that, even in difficult cases involving competing instabilities, it
may be possible to find predictors for $T_{\rm C}$ associated with
simple properties of the ground state. The possibility of extending
such a conclusion to other materials with competing instabilities
remains to be confirmed.

\subsection{Additional remarks}

Abrahams {\sl et al}.\cite{abrahams68} noted that the validity of
their simple empirical rule implies that very different materials must
present similar properties of some sort. Indeed, as discussed
theoretically by Lines,\cite{lines69,lines-book1977} the applicability
of Eq.~(\ref{eq:empirical}) to a set of diverse ferroelectrics
indicates that there exists an effective force constant (in essence,
this would be the proportionality constant between $T_{\rm C}$ and
$u_{gs}^{2}$) that (1) is probably dominated by long-range
dipole-dipole interactions and (2) is quantitatively similar for all
the materials considered in Ref.~\onlinecite{abrahams68}. This
interpretation is consistent with our results: We have found that
variations in $C_{\rm FE}$ have a large impact on $T_{\rm C}$; hence,
if a $T_{\rm C}$ predictor that disregards such variations applies to
a set of materials, it follows that $C_{\rm FE}$ must be similar for
all of them. We believe that, as long as we are dealing with {\em
  simple} cases (e.g., in absence of a materials-dependent competition
between structural instabilities), such reasonings probably
apply. Yet, in view of the above-described subtleties associated with
the Pb-based relaxors, it is legitimate to wonder how many intricate
and material-specific behaviors are hiding behind the result of
Abrahams {\sl et al}., and how much their empirical rule really tells
us about the nature of the interatomic interactions in each of the
specific compounds they considered.

In the same spirit, Grinberg and Rappe\cite{grinberg04} suggested that
the PTO-based solid solutions they investigated must present somewhat
similar Landau potentials. More specifically, they worked with a
Landau energy of the form
\begin{equation}
F = \alpha (T-T_{\rm C}) P^{2} + \beta P^{4} \, ,
\label{eq:landau}
\end{equation}
with $\alpha, \beta > 0$, to justify the relationship
\begin{equation}
T_{\rm C} = 2\beta P_{gs}^{2}/\alpha \, , 
\label{eq:tc-landau}
\end{equation}
and inferred that the compounds they studied (for which $T_{\rm
  C}/P_{gs}^{2}$ is approximately constant) must present comparable
$\beta/\alpha$ ratios. This is a tempting interpretation that seems to
give us some physical insight into the energetics of the polar
instabilities in these materials. However, noting that the findings of
these authors on Pb-based relaxors probably rely on subtle effects
involving competing instabilities, and that such effects cannot be
modeled within a simple Landau scheme, we should be careful to avoid
overinterpreting such observations.

Finally, let us note that we have limited our discussion to $T_{\rm
  C}$ predictors that have a clear justification, may it be empirical,
theoretical, or heuristic. We have purposely avoided the consideration
of other possibilities with a less clear basis. For example, it may be
tempting to consider the Landau potential of Eq.~(\ref{eq:landau}) and
derive possible rules like $T_{\rm C} = 2\beta P_{gs}^{2}/\alpha$
[Eq.~(\ref{eq:tc-landau})] or $T_{\rm C}^{2} = 4\beta
|E_{gs}|/\alpha^{2}$. However, while the former seems equivalent to
Eq.~(\ref{eq:empirical}) for $n = 2$, and the latter may resemble
Eq.~(\ref{eq:energy}), it must be emphasized that these identities
cannot be used to justify a predictor for $T_{\rm C}$. The reason is
that it is perfectly legitimate to choose $T_{\rm C}$, $P_{gs}$, and
$E_{gs}$ as independent parameters of the Landau potential and, hence,
no relationship among these quantities needs to hold. Note, for
example, that in addition to the two expressions just mentioned, we
may write others such as $T_{\rm C} = 2 |E_{gs}|/(\alpha P_{gs}^{2})$,
which renders a very different and equally unjustified relationship
between our properties of interest.

\section{Conclusions}

We have examined an effective model for PbTiO$_{3}$, a material with a
representative soft mode-driven structural transition, to investigate
which features of the potential control the transition temperature
$T_{\rm C}$. Our main result is that $T_{\rm C}$ correlates strongly
with the energy difference between the high-symmetry and low-symmetry
structures ($E_{gs}$). In contrast, we find that the magnitude of the
symmetry-breaking distortion ($u_{gs}$) is a less robust predictor,
although it can be expected to work well in typical cases in which
$|E_{gs}|$ and $|u_{gs}|$ are strongly correlated. Additionally, our
results reveal the sizable impact that subtle features of the energy
surface have on the computed $T_{\rm C}$, providing us with useful
information for the construction of more accurate model potentials
from first principles.

By comparing our results with existing literature, we can conclude
that: (1) Whenever simple $T_{\rm C}$ predictors work well for a
family of materials, this is indicative that the potentials of such
compounds share some common features. This conclusion is in agreement
with previous observations by other
authors.\cite{abrahams68,lines-book1977} (2) Whenever the simple
predictors fail, this suggests the occurrence of subtle structural and
lattice-dynamical effects involving strong anharmonic interactions
between modes.

We thus hope our results will bring new insights to the analysis of
complex phase-transition and lattice-dynamical phenomena, and permit
more effective computational works to design materials with tailored
temperature-dependent properties.

This work was supported by MINECO-Spain (Grants No. MAT2010-18113 and
No. CSD2007-00041) and CSIC [JAE-doc program (JCW)].

\appendix*

\section{Fitting the $P(T)$ curves}

To analyze our data and determine $T_{\rm C}$ for each considered
model in a robust and reliable way, we employed a fitting procedure
that assumes a heuristic form for $P(T)$. More precisely, we used
\begin{equation}
P(T) = \left\{ 
\begin{array}{ll}
\mu(T_{\rm C}-T)^{\delta} & {\rm for}\;\; T<T_\mathrm{C} \\ 0 & {\rm
  for}\;\; T>T_{\rm C} \\
\end{array} \; ,
\right.
\end{equation}
which depends on the three free parameters $\mu$, $T_{\rm C}$, and
$\delta$. This functional form is compatible with the description of a
second order phase transition (which corresponds to $\delta = 1/2$
within Landau theory), and is also flexible enough to capture more
complex behaviors appearing when the transition is discontinuous. (We
numerically found that $\delta \simeq 1/5$ reproduces well the results
for our reference model, as well as the typical first-order transition
described by a sixth-order Landau theory.) The good quality of the
fits can be appreciated in Fig.~\ref{fig1}(c), which shows a
representative case.

Additionally, we introduced a linear dependence of the free parameters
-- $T_{\rm C}$, $\mu$, and $\delta$ -- on the specific tuned
properties (i.e., $E_{gs}/E^0_{gs}$ or $P_{gs}/P^0_{gs}$) to be able
to plot the surfaces appearing in Fig.~\ref{fig3}. There we also
report the deviation (``fit error'') between the fitted curves and the
computed polarization values, which turns out to be very small except
in the immediate vicinity of $T_{\rm C}$.


\begin{thebibliography}{21}
\bibitem [{fn-()}]{fn-hysteresis}%
  \BibitemOpen
  \href@noop {} {}\bibinfo {note} {The computed hysteresis is expected to
  depend strongly on the size of the simulated supercell. In our case, the
  supercell is probably too small to render a realistic description of the
  hystereis in the ferroelectric transition of PTO, as we get a very small
  effect.}\BibitemShut {Stop}%
\end{thebibliography}
\end{document}